\documentclass{article}

\usepackage{arxiv}

\usepackage[utf8]{inputenc} 
\usepackage[T1]{fontenc}    
\usepackage{hyperref}       
\usepackage{url}            
\usepackage{booktabs}       
\usepackage{amsfonts}       
\usepackage{amsmath}       
\usepackage{nicefrac}       
\usepackage{microtype}      
\usepackage{lipsum}		
\usepackage{graphicx}
\usepackage{natbib}
\usepackage{doi}
\usepackage{caption}
\usepackage{subcaption}

\title{High-energy X-ray spectrum reconstruction: solving the inverse problem from optimized multi-material transmission measurements}

\author{ \href{https://orcid.org/0000-0000-0000-0000}{\includegraphics[scale=0.06]{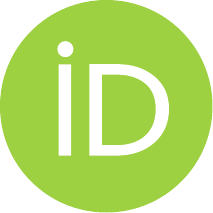}\hspace{1mm}Arthur Walker} \\
	Commissariat à l'Energie Atomique,\\
	Chemin du Ru,\\
	91680 Bruyères-le-Châtel \\
	\texttt{arthur17.walker@gmail.com} \\
        \And
        \href{https://orcid.org/0000-0000-0000-0000}{\includegraphics[scale=0.06]{orcid.pdf}\hspace{1mm}Alexandre Friou} \\
	Commissariat à l'Energie Atomique,\\
	Chemin du Ru,\\
	91680 Bruyères-le-Châtel \\
	\texttt{alexandre.friou@cea.fr} \\
	\And
        \href{https://orcid.org/0000-0000-0000-0000}{\includegraphics[scale=0.06]{orcid.pdf}\hspace{1mm}Kévin Ginsburger} \\
	Commissariat à l'Energie Atomique,\\
	Chemin du Ru,\\
	91680 Bruyères-le-Châtel \\
	\texttt{kginsburger@gmail.com} \\
}

\renewcommand{\shorttitle}{High-energy X-ray spectrum reconstruction}

\hypersetup{
pdftitle={High-energy X-ray spectrum reconstruction},
pdfauthor={Arthur Walker, Alexandre Friou, Kevin Ginsburger},
pdfkeywords={spectrum estimation, flash X-ray, transmission measurement, genetic algorithm, ill-posed inverse problem},
}

\begin{document}
\maketitle

\begin{abstract}
Reconstructing the unknown spectrum of a given X-ray source is a common problem in a wide range of X-ray imaging tasks. For high-energy sources, transmission measurements are mostly used to recover the X-ray spectrum, as a solution to an inverse problem. While this inverse problem is usually under-determined, ill-posedness can be reduced by improving the choice of transmission measurements. A recently proposed approach optimizes custom thicknesses of calibration materials used to generate transmission measurements, employing a genetic algorithm to minimize the condition number of the system matrix before inversion.

In this paper, we generalize the proposed approach to multiple calibration materials and show a much larger decrease of the condition number of the system matrix than thickness-only optimization.
Additionally, the spectrum reconstruction pipeline is tested in a simulation study with a challenging high-energy Bremsstrahlung X-ray  source encountered in Linear Induction Accelerators (LIA), with strong scatter noise. Using this approach, a realistic noise level is obtained on measurements. A generic anti-scatter grid is designed to reduce noise to an acceptable -yet still high- noise range. A novel noise-robust reconstruction method is then presented, which shows much less sensitive to initialization than common expectation-maximization approaches, enables a precise choice of spectrum resolution and a controlled injection of prior knowledge of the X-ray spectrum. 

\end{abstract}


\section{INTRODUCTION}

The reconstruction of the unknown spectrum of a given X-ray source is a common problem in a wide range of X-ray imaging tasks~\citep{yaffe_application_2004, duan_ct_2011}. If the source flux is low, spectrometers are usually preferred to estimate the spectrum. If a very precise modeling of the source is available, a good knowledge of the X-ray spectrum can be obtained, provided that the model input parameters, such as voltage, are measured precisely enough during the pulse~\citep{wood_shot-by-shot_2018}.
When none of the two previous conditions are met, transmission measurements are most frequently used to recover the X-ray spectrum, as a solution to an inverse problem~\citep{waggener_x-ray_1999, armbruster_spectrum_2004, paniak_enhanced_2005, sidky_robust_2005}.

This inverse problem is usually under-determined, because a high resolution of the reconstructed spectrum is required. It is also ill-conditioned, making the spectral estimation unstable and very sensitive to noise. The ill-posedness of this inverse problem can be reduced using a parametric model of the reconstructed spectrum~\citep{zhao_indirect_2015, fitzgerald_semiempirical_2021}. However, these model-based methods restrict, by design, the space of possible solutions, thus requiring a fine and general enough physical modelling of the spectrum prior to reconstruction.

Another way to reduce ill-posedness is to improve the quality of the set of transmission measurements. In particular, the recent approach described in~\citep{li_em_2021} proposed to compute custom thicknesses of calibration materials used to generate transmission measurements, by optimizing on the condition number of the system matrix used for inversion. Using a genetic algorithm, the interest of optimized measurements to reconstruct spectra was demonstrated, in comparison with common linear slab phantoms. 

While the approach proved to be very efficient for the configurations tested in~\citep{li_em_2021}, their simulation study was restricted to unrealistically small amounts of Poisson noise, and relatively low-energy spectra. In this work, a challenging high-energy Schiff spectrum~\citep{schiff_energy-angle_1951} is used to simulate transmission measurements using the Monte-Carlo N-Particle code (MCNP4C) with realistic noise levels. The Schiff spectrum is a typical model for thin-target Bremsstrahlung spectra encountered in radiographic sources based on Linear Induction Accelerators (LIA), used to perform high-energy flash X-ray imaging ~\citep{estre_high-energy_2013}. We show that the method presented in~\citep{li_em_2021} is not readily applicable to this real-world reconstruction problem. Three  improvements are thus proposed to obtain a robust spectrum reconstruction. Firstly, the transmission measurement optimization, reduced to variations of material thicknesses in~\citep{li_em_2021}, is extended to multiple materials by modifying the genetic algorithm. We show that using multiple materials yields a much larger decrease of the condition number of the system matrix than thickness-only optimization. Secondly, instead of a generic Poisson noise, a realistic simulated scatter noise is used to evaluate the ability to recover the spectrum. Using this approach, we observe a much higher noise level on measurements, which makes the design of a noise reduction setup mandatory. As such, an anti-scatter grid is proposed, reducing noise to an acceptable range for spectrum reconstruction, but still much higher than noise levels encountered in~\citep{li_em_2021}.  Thirdly, once measurements are optimized and the experimental setup is fixed, a novel reconstruction method is presented, which shows much less sensitive to initialization than expectation-maximization approaches~\citep{li_em_2021}, enables a precise choice of spectrum resolution and a controlled injection of prior knowledge of the X-ray spectrum. 

\section{METHODS}

\subsection{Transmission measurement model}
\label{transmission_measurement_model}

\begin{figure}[h!]
\begin{center}
\includegraphics[width=10cm]{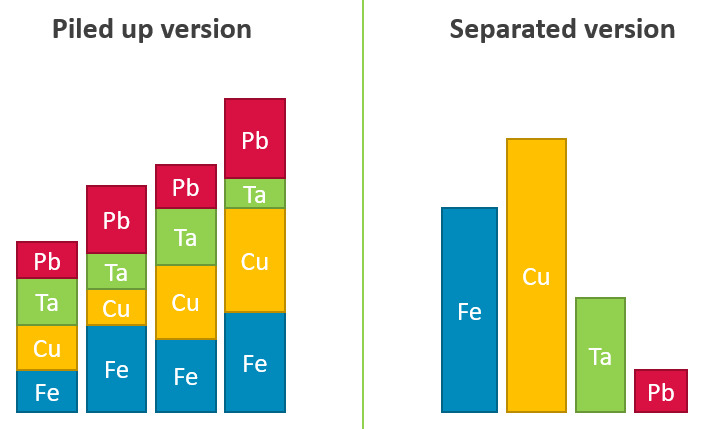}
\end{center}
\caption{Illustration of the two configuration considered for 4 measurement slabs}
\label{fig:algorithm_versions}
\end{figure}

As illustrated in figure~\ref{fig:algorithm_versions}, two types of transmission measurements are considered, which correspond to the two experimental configurations tested in this study. 

In what follows, the detector spectral response $D(E)$ is not accounted for and taken as unity. For each measurement, the forward transmission model is given by the Beer-Lambert law. Given an X-ray source with spectrum $S(E)$, the transmitted intensity $I$ of a measurement writes

\begin{equation}
I = \int_{E} S(E)D(E)e^{-l\mu(E)} \,dE
\label{eq:01}
\end{equation}   

where $l$ is the thickness and $\mu(E)$ the linear attenuation coefficient of the material used for this measurement.

\subsubsection{One material per measurement}

This configuration corresponds to the right side of figure~\ref{fig:algorithm_versions}.
For $M$ measurements $y_i$ and with an even discretization of the spectrum into $N$ bins, the forward problem can be cast into a set of linear equations

\begin{equation}
y_{i} = \sum_{j=1}^{N} e^{-l_{i} \mu_{i,j}} s_{j}, \;i \in [\![ 1, M ]\!], \;j \in [\![ 1, N ]\!]
\label{eq:02}
\end{equation} 

where $l_i$ is the thickness of the $i$-th measurement and $\mu_{i,j}$ is the linear attenuation coefficient value for the $i$-th measurement at the discrete energy level $j$.

Equation~\ref{eq:02} is conveniently put in the matrix form as

\begin{equation}
\textbf{y} = A \times \textbf{s}
\label{eq:03}
\end{equation} 

with $\textbf{y} = (y_{i}) \in \mathbb{R}^{M}$ the vector of transmission measurements, $\textbf{s} = (s_{j})\in \mathbb{R}^{N}$ the vector of the discretized spectrum and $A = (e^{-l_{i}\mu_{i,j}}) \in \mathbb{R}^{M \times N}$ the forward system matrix.

\subsubsection{Multiple materials per measurement}

This configuration is illustrated at the left side of figure~\ref{fig:algorithm_versions}. Each measurement contains $K$ layers of distinct materials. 
Without any loss of generality, we consider that each measurement contains the same number of layers with the same material order. Only the thickness of each material layer is changed, and can be set to zero. With this assumption, the attenuation coefficient  no longer depends on the measurement number $i$. The transmitted intensity $y_i$ for the $i$-th measurement thus writes

\begin{equation}
y_i = \sum_{j=1}^{N} \prod_{k=1}^{K} e^{-l_{i,k} \mu_{k,j}} s_{j},  \;i \in [\![ 1, M ]\!], \;j \in [\![ 1, N ]\!], \;k \in [\![ 1, K ]\!]
\label{eq:04}
\end{equation}

where $l_{i,k}$ is the thickness of the $k$-th layer of the $i$-th measurement and $\mu_{k,j}(E)$ is the linear attenuation coefficient of the material $k$ at energy level $j$. The forward system matrix in equation~\ref{eq:03} is modified as $A = (e^{-l_{i,k}\mu_{k,j}}) \in \mathbb{R}^{M \times N}$.

\subsection{Multi-material thickness optimization}

Using multiple materials in transmission measurements implies some modifications  to the thickness optimization genetic algorithm described in ~\citep{li_em_2021}. Two optimizations with slightly different implementations, corresponding to the two types of transmission measurements discussed above, are here considered and tested.


\subsubsection{Multiple materials per measurement}
\label{multiple_materials_per_measurement}

When multiple piled up materials are allowed for each measurement, the genetic algorithm optimizes on two-dimensional matrices of size $M \times K$ instead of one-dimensional vectors in the single material case. Each cell of the matrices contains the current thickness of the column's corresponding material for the line's corresponding measurement. In comparison with~\citep{li_em_2021}, the main steps of the genetic optimization are modified as follows:

\begin{itemize}
    \item \textbf{Initialization} : $N_{pop}$ matrices are initialized with a randomly chosen value in each cell, such that the sum on each row (i.e. the total thickness of the corresponding measurement slab) remains in a chosen interval $[l_{min}, l_{max}]$
    \item \textbf{Crossover} : the crossover formula used in~\citep{li_em_2021} is applied similarly to the rows of the system matrix
    \item \textbf{Mutation} : instead of simply modifying the value of the mutated cell as in~\citep{li_em_2021}, the sum of values on the row, corresponding to the total thickness, is modified as well as the proportion of materials
    \end{itemize}

\subsubsection{One material per measurement}
\label{one_material_per_measurement}

When only one material is allowed per measurement, two-dimensional $M \times K$ matrices are also used but with only one non-zero value per row, at the column corresponding to the employed material. The main steps of the genetic optimization are modified as follows:
\begin{itemize}
    \item \textbf{Initialization} : $N_{pop}$ matrices are initialized with a randomly chosen value in only one cell per row, and the other cells are initialized to zero
    \item \textbf{Crossover} : If the two parents are using the same material, the situation refers to the case of ~\citep{li_em_2021}. Otherwise, the two children receive one material each, with thickness values crossed with the same formula
    \item \textbf{Mutation} : Each mutation on a row also has a probability of changing the material used for the corresponding measurement
    \end{itemize}

\subsection{Geometry design and simulation}

A basic MCNP4C geometry was built for each set of measurements. Transmission measurement phantoms were represented as cylinders of known thicknesses and radii. Cylinders axes are aiming at the photon source, which is modeled as a point source located 180 cm away from the phantoms. A Schiff spectrum, corresponding to a maximum electron energy of 20 MeV, incident on a 1.2 mm thick tantalum target, is used. The photons are emitted within a cone with a constant angular distribution, chosen in order to cover all the measurement devices. Both photon and electron interactions are modeled, so as to accurately account for scattering. The filling medium is air. All these parameters were chosen to be as close as possible to reality.

The detectors in the simulation are modeled as simple fluence tallies, placed behind each measurement slab. Consistent modeling of detectors in the simulation and accounting for their spectral responses in the optimization process, is left to future work.

As mentioned earlier, the MCNP4C simulation accounts for the presence of scatter noise in the measurements, produced during the passage of X-rays through materials. A specific simulation setup was designed to reduce this scatter noise and obtain clean enough measurements for the spectrum reconstruction. 

The first approach employed to decrease scatter noise was to optimize the placement of the measurement slabs within the experimental setup and space them out in order to reduce the interaction between  particles coming through the materials. The differential evolution algorithm \cite{storn1997differential} was used to compute the optimal placement of $M$ points constrained in a circle by minimizing the electrostatic potential between them. With this method, the scatter noise was reduced by half compared to slabs placed on a circle. While significant, this reduction was not sufficient to exploit transmission measurements for spectrum reconstruction. To further reduce this noise, a specific anti-scatter grid was designed. This grid consists of a 20 cm deep lead layer between the measurement slabs and the dose sensors. Cylinders of small radius are carved in the lead behind each measurement slab in order to absorb all particles except source photons, yielding the expected signal. Using this anti-scatter grid reduces the scatter noise to an exploitable amount of around $3\%$, allowing the measurements to be used for the spectrum reconstruction.

\subsection{Reconstruction algorithm}

\subsubsection{Adaptive resampling based on a typical spectrum}

\begin{figure}[h!]
\begin{center}
\includegraphics[width=10cm]{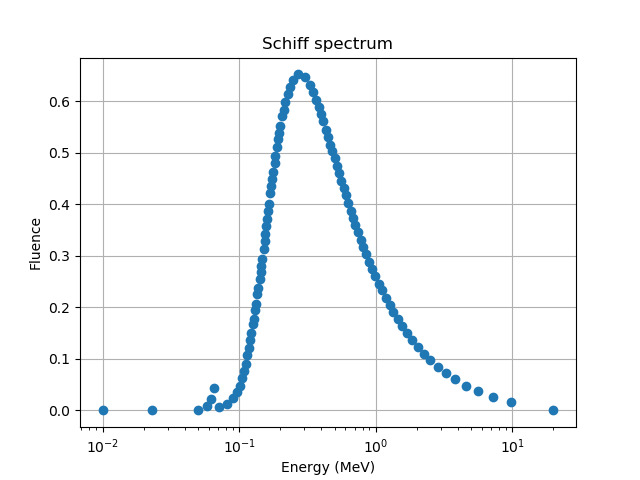}
\end{center}
\caption{Typical Schiff spectrum optimally discretized to $N = 100$ energy points}
\label{fig:schiff_spectrum_better_discretized}
\end{figure}

Prior to the reconstruction algorithm itself, a preliminary step of dimension reduction is realized on a typical spectrum presenting roughly the same characteristics as the unknown spectrum. As shown in figure~\ref{fig:schiff_spectrum_better_discretized}, by sampling uniformly from the cumulative integral of this typical spectrum's derivative, an approximately optimal choice of the $N$ energy bins is obtained, allowing an efficient spectrum representation, later employed during the optimization process of the reconstruction.

\subsubsection{Spectrum reconstruction as an interpolation}

The spectrum is reconstructed on the optimal energy sampling presented above, using experimental or simulated measurements $\textbf{y}$. A candidate spectrum $\textbf{s}_{cand}$ is initialized and then modified during the optimization process. The measurements computed through the forward model $\textbf{y}_{cand} = A \times \textbf{s}_{cand}$ are expected to be as close to $\textbf{y}$ as possible.

The optimization problem for the spectrum reconstruction thus writes :
\begin{equation}
    \underset{\textbf{s}_{cand}}{\mathrm{argmin}} \quad \lVert \textbf{y} - A \times 
    \textbf{s}_{cand} \rVert^2
\end{equation}

A straightforward method would consist in performing a spectrum discretization over the optimal energy sampling values $E_1, ... ,E_N$ where $N$ is usually large to obtain a good resolution of the spectrum. However, a large $N$ increases the degrees of freedom and makes the optimization process harder. It is therefore necessary to reduce the number of parameters used to describe the spectrum. The method chosen here is to sample the spectrum with a small number $P$ of interpolation points. This requires the spectrum to be continuous and deprived of high variation peaks, which is a general feature of high-energy Bremsstrahlung spectra.

\begin{figure}[h!]
\begin{center}
\includegraphics[width=10cm]{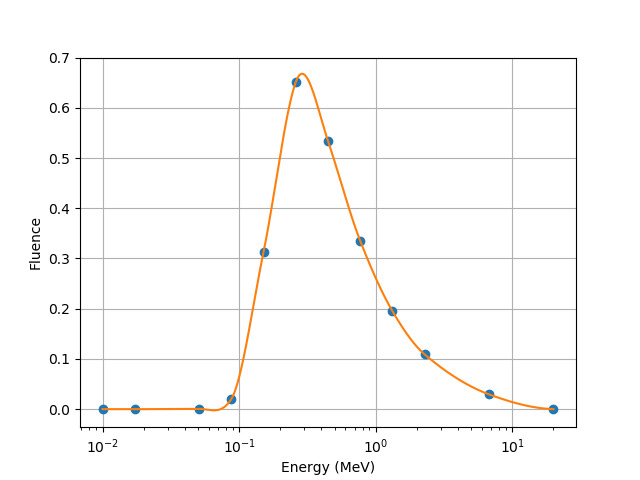}
\end{center}
\caption{Schiff spectrum interpolated by P = 12 interpolation points distributed in the energy interval}
\label{fig:schiff_interpolated}
\end{figure}

As shown in figure~\ref{fig:schiff_interpolated}, the spectrum is thus fully described by $P$ interpolation points, which coordinates can evolve both in energy and magnitude (e.g. $2P$ degrees of freedom). At each optimization step, the points are interpolated by a piecewise cubic polynomial function \cite{akima1970new}, then projected over the $N$ energy intervals. $\textbf{y}_{cand}$ is then computed and new values of coordinates for the interpolation points can be determined.

\subsubsection{Reconstruction algorithm}

A trust-region algorithm for constrained optimization \cite{conn2000trust} is used to compute the interpolation points corresponding to the approximated spectrum. Minimum and maximum energy levels are enforced, based on a known low energy cut-off and the maximum energy of electrons. The spectrum magnitude is constrained using a simple step function. In addition to this restriction, the spectrum is normalized by its integral during each step of the optimization process.

For this algorithm, the $P$ abscissa of the interpolation points are fixed, which improves optimization performances and decreases execution time. However, other algorithms might be more efficient with the abscissa taken as additional degrees of freedom. In the case of this study, these abscissa values are taken as evenly spaced inside the energy interval, in logarithmic scale. The optimization algorithm is then applied to the $P$ magnitudes of the interpolation points, with a fixed number of iterations.

\section{RESULTS}

\subsection{Impact of multiple materials}
\label{multiple_materials_impact}

A comparison between our multi-material genetic algorithm and the version implemented in~\citep{li_em_2021} was performed using the same setup. A number of $M = 8$ measurements was fixed, $N=100$ energy bins and $K=4$ materials (iron, copper, tantalum and lead) were used. For a fair comparison, all other hyper-parameters, including the number of generations, the population size, the crossover and mutation probabilities, the target distribution mean and the best fitness boosting factor, were kept identical to~\citep{li_em_2021}.

\subsubsection{Algorithms performance}

\begin{figure}[h!]
    \centering
    \begin{subfigure}[b]{0.4\textwidth}
        \centering
        \includegraphics[width=\linewidth]{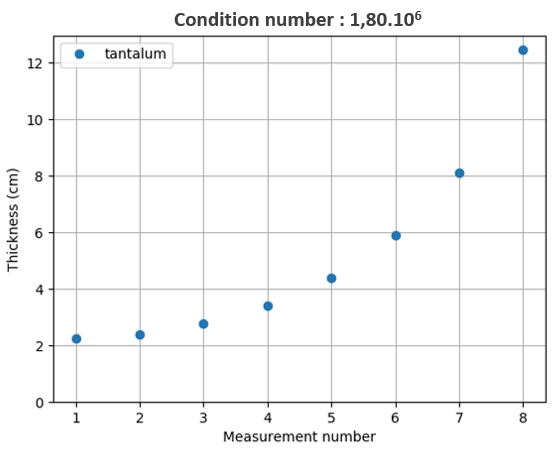}
        \caption{Original single-material version of the genetic algorithm}
        \label{fig:results_1mat}
    \end{subfigure}  
    \hfill
    \begin{subfigure}[b]{0.4\textwidth}
        \centering
        \includegraphics[width=\linewidth]{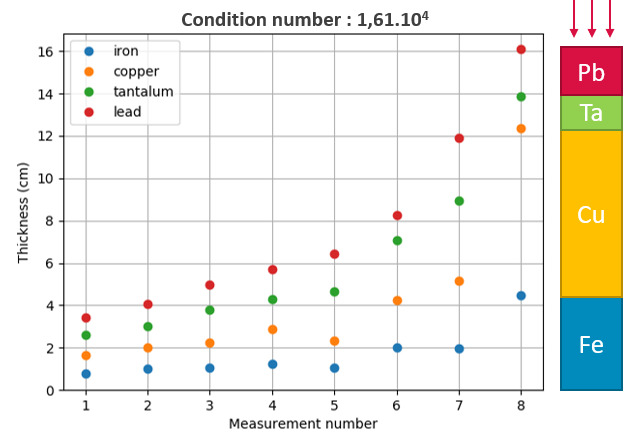}
        \caption{Version with multiple materials allowed per measurement, graphic representation of the optimized slab for measurement 8}
        \label{fig:results_empilement}
    \end{subfigure}
    \hfill
    \begin{subfigure}[b]{0.4\textwidth}
        \centering
        \includegraphics[width=\linewidth]{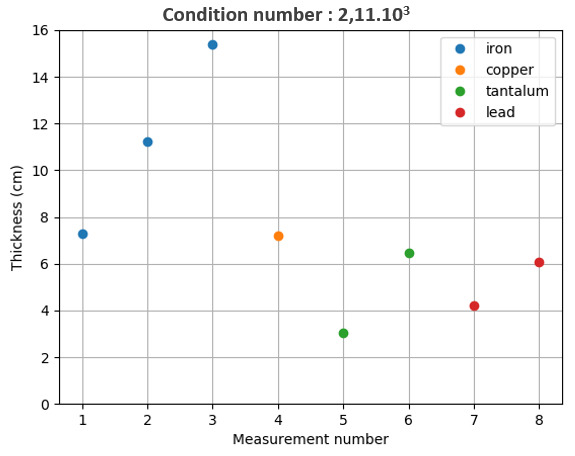}
        \caption{Version with only one material allowed per measurement, ordered by material}
        \label{fig:results_separation}
    \end{subfigure}
    \caption{Optimal arrangement of thicknesses and achieved condition number for the different versions of the genetic algorithm}
    \label{fig:results_first_optim}
\end{figure}

The work done in~\citep{li_em_2021} led to a great reduction in orders of magnitude of the forward matrix condition number, and highlighted an exponential distribution of thicknesses for the optimal measurement set using one material. Using this one-material genetic algorithm, the optimal arrangement of thicknesses plotted on figure \ref{fig:results_1mat} was obtained, with a condition number of $1.8\times 10^6$.

Figure \ref{fig:results_empilement} displays the results obtained using the first type of transmission measurements with several materials, described in section~\ref{multiple_materials_per_measurement}. When multiple piled up materials are allowed per measurement, a significant improvement of the previous results is observed, with a further reduced condition number of $1.61\times 10^4$.

Figure \ref{fig:results_separation} shows the results obtained using a single material per measurement.
With this second version of the algorithm presented in \ref{one_material_per_measurement}, an even greater improvement is observed with a highly reduced condition number of $2.11\times 10^3$.

\subsubsection{Stability of the optimization}

Regardless of the performance of the different algorithms, another main goal of the proposed algorithms is to ensure a good stability of the optimized measurements. To compare the stability of the multi-material algorithms developed in this study, both were executed several times with the same parameters. The set of materials obtained for each optimization was then plotted.

\begin{figure}[h!]
    \centering
    \begin{subfigure}[b]{0.7\textwidth}
        \centering
        \includegraphics[width=\linewidth]{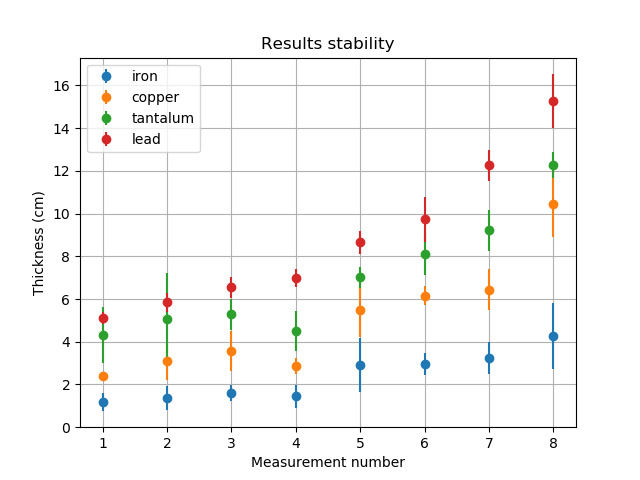}
        \caption{Mean optimal thicknesses and standard deviation for 20 optimizations with the same parameters when multiple materials are allowed per measurement}
        \label{fig:piled_up_stability}
    \end{subfigure}  
   
    \begin{subfigure}[b]{0.7\textwidth}
        \centering
        \includegraphics[width=\linewidth]{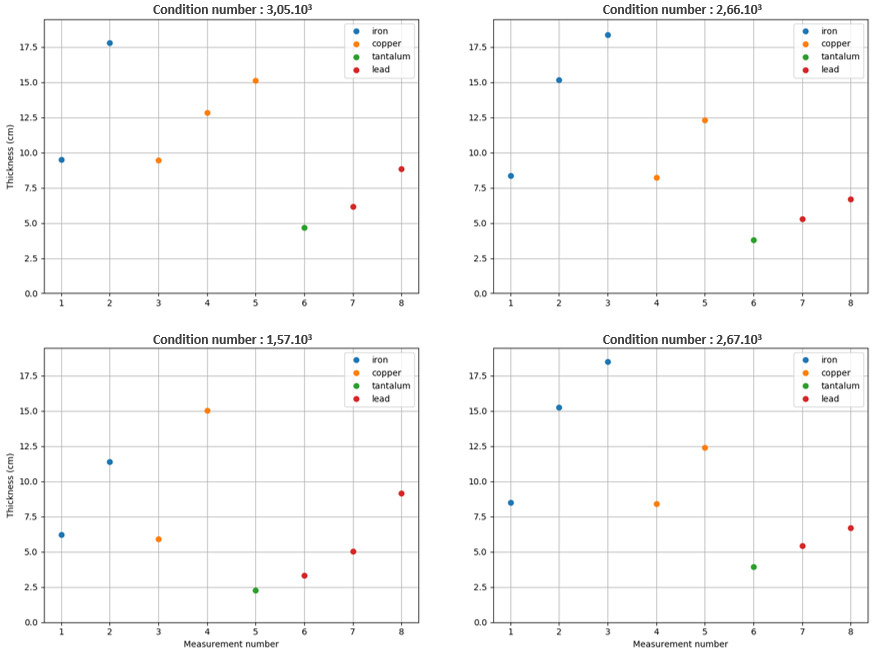}
        \caption{Optimal arrangement of thicknesses for 4 optimizations with similar parameters when only one material is allowed per measurement}
        \label{fig:separated_stability}
    \end{subfigure}

    \caption{Stability study for both versions of the multi-material genetic algorithm}
    \label{fig:stability_first_optim}
\end{figure}

Results are shown in figure~\ref{fig:stability_first_optim}. In figure~\ref{fig:piled_up_stability} where multiple materials per measurement are allowed, the mean thickness value and the standard deviation of each piled up material for each measurement are plotted. However, in figure~\ref{fig:separated_stability} where only one material is allowed per measurement, the plots have to be separated because a given measurement can correspond to different materials for different optimizations.

The result of the stability comparison is clear. As illustrated on figure~\ref{fig:piled_up_stability}, the algorithm version with piled up materials is quite unstable, with a significant standard deviation on material thicknesses between experiments and a strong variance on the condition number. Conversely, for the algorithm with only one material per measurement, the stability is satisfying. As shown in figure~\ref{fig:separated_stability}, each measurement is attributed almost always the same material with the same thickness across experiments. 
The reduction of the research space size thus plays a key role in the stability improvement.

\subsection{Reduction of scatter noise}

\begin{figure}[h!]
    \centering
    \begin{subfigure}[b]{0.5\textwidth}
        \centering
        \includegraphics[width=\linewidth]{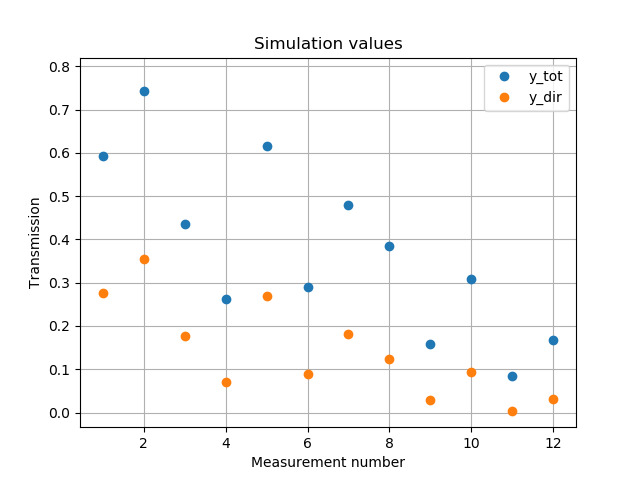}
        \caption{Case of naive configuration (config. 1)}
        \label{fig:barillet_tot_dir}
    \end{subfigure}  
    \hfill   
    \begin{subfigure}[b]{0.5\textwidth}
        \centering
        \includegraphics[width=\linewidth]{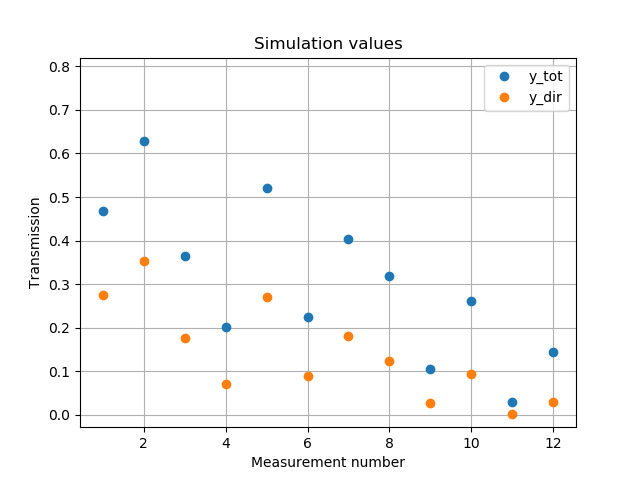}
        \caption{Case of optimally distributed slabs (config. 2)}
        \label{fig:hedgehog_tot_dir}
    \end{subfigure}    
    \hfill   
    \begin{subfigure}[b]{0.5\textwidth}
        \centering
        \includegraphics[width=\linewidth]{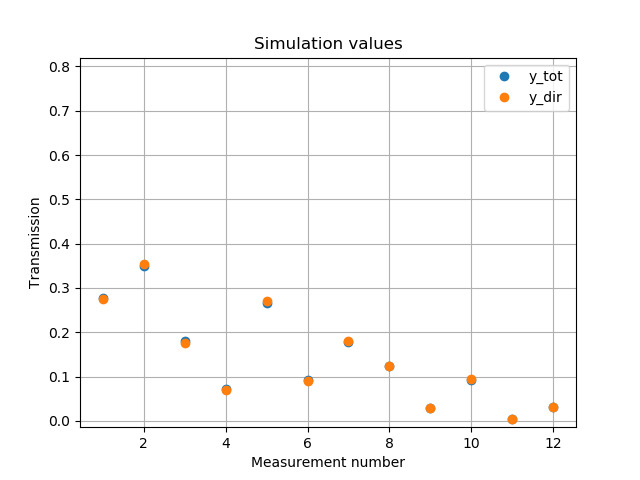}
        \caption{Case of optimally distributed slabs with a lead layer (config. 3)}
        \label{fig:hedgehog_shield_tot_dir}
    \end{subfigure}    

    \caption{Total ($\textrm{y}_{tot}$) and direct ($\textrm{y}_{dir}$) measured rays for the 3 tested experimental configurations}
    \label{fig:configuration_noises}
\end{figure}

Different geometric configurations have been tested in MCNP4C to reduce scatter noise while keeping an easy-to-design setup. For each of them, a simulation was run using the measurement slabs returned by the one-material-per-measurement version of the genetic algorithm for $M = 12$ measurements. In this study we consider that scattered rays represent the entire measurement noise, which is a reasonable approximation. The Monte-Carlo simulation code allows the calculation of the theoretical unscattered rays along with the measurement of total (unscattered and scattered) rays. The objective is to design a configuration in which the measured total rays are as close to the theoretical direct rays as possible. Total and unscattered rays have been measured for each simulation and are plotted in figure~\ref{fig:configuration_noises}.

Three configurations were tested : a straightforward design where measurement slabs were placed on the edge of a circle of given radius (config. 1, figure~\ref{fig:barillet_tot_dir}), a reworked configuration in which slabs were optimally distributed in a circle (config. 2, figure~\ref{fig:hedgehog_tot_dir}), and a last design where a thick lead anti-scatter grid was added between the measurement slabs and the sensors of configuration 2 (config. 3, figure~\ref{fig:hedgehog_shield_tot_dir}).

\subsection{Schiff spectrum reconstruction}

\subsubsection{Nominal configuration}

\begin{figure}[h!]
\begin{center}
\includegraphics[width=10cm]{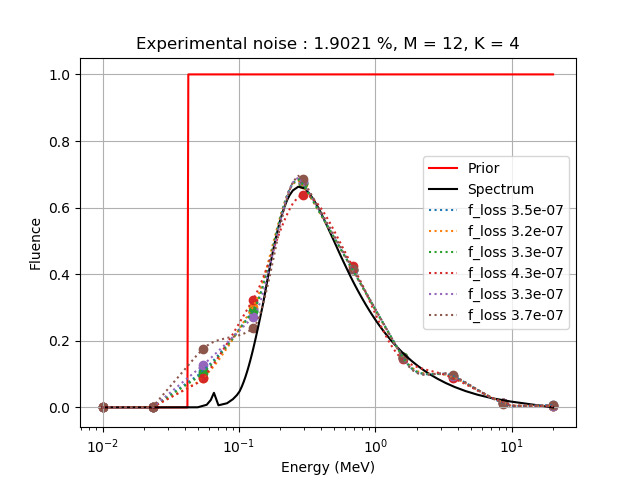}
\end{center}
\caption{Spectrum reconstruction in the nominal configuration}
\label{fig:nominal_reconstruction}
\end{figure}

In this study, the nominal configuration for the spectrum reconstruction consists in : $M = 12$ measurements, $N = 100$ energy intervals, $K = 4$ different materials (iron, copper, tantalum and lead), only one material allowed per measurement in the genetic measurement set optimization (hyperparameters : 500 generations, 1000 in population size), config. 3 for the simulation configuration, and $P = 10$ interpolation points for the reconstruction algorithm.

For this nominal configuration, the reconstruction algorithm has been applied multiple times. Reconstructed spectrums (dotted lines) are plotted on figure~\ref{fig:nominal_reconstruction}, along with the objective theoretical spectrum (solid black line).

\subsubsection{Ablation studies}

\begin{figure}[h!]
    \centering
    \begin{subfigure}[b]{0.5\textwidth}
        \centering
        \includegraphics[width=\linewidth]{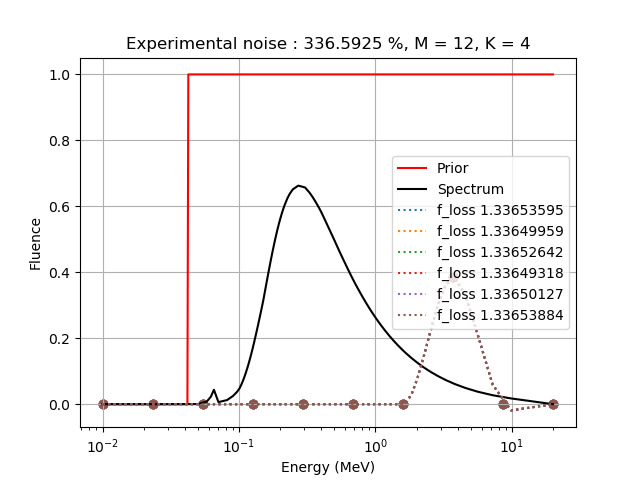}
        \caption{Reconstruction with no anti-scatter grid used}
        \label{fig:no_anti-scatter_grid_reconstruction}
    \end{subfigure} 
    \hfill   
    \begin{subfigure}[b]{0.5\textwidth}
        \centering 
        \includegraphics[width=\linewidth]{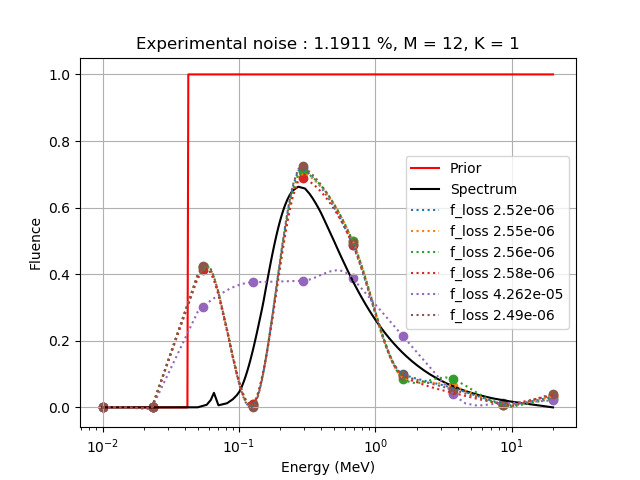}
        \caption{Reconstruction with only one material}
        \label{fig:1_mat_reconstruction}
    \end{subfigure} 
    \hfill   
    \begin{subfigure}[b]{0.5\textwidth}
	\centering
        \includegraphics[width=\linewidth]{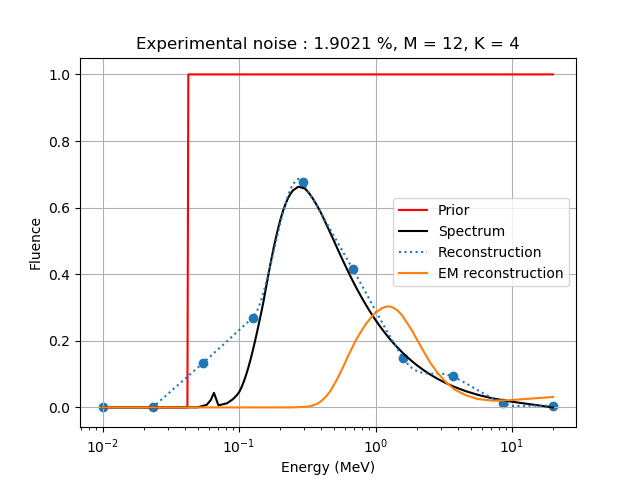}
        \caption{Reconstrucion with expectation-maximization algorithm (EM reconstruction) compared to the algorithm described above (Reconstruction)}
        \label{fig:EM_reconstruction}
    \end{subfigure} 

    \caption{Ablation study of the spectrum reconstruction in nominal configuration}
    \label{fig:ablation_study}
\end{figure}

Finally, ablation studies were performed to evaluate the influence of every part of the spectrum estimation pipeline on the reconstruction accuracy. Figure~\ref{fig:ablation_study} highlights the relative improvements obtained with each major step of the reconstruction method.

More precisely, reconstructions have been performed independently in the nominal configuration in the cases where :
\begin{itemize}
    \item No anti-scatter grid was used for the measurements (figure~\ref{fig:no_anti-scatter_grid_reconstruction})
    \item Only one material (iron) was used for the experimental slabs (figure~\ref{fig:1_mat_reconstruction})
    \item The expectation-maximization (EM) algorithm was used for the reconstruction instead of the algorithm proposed in this study (figure~\ref{fig:EM_reconstruction})
\end{itemize}

\section{DISCUSSION}

\subsection{Impact of multiple materials}

The work done in~\citep{li_em_2021} led to a great reduction in orders of magnitude of the forward matrix condition number, and highlighted an exponential distribution of thicknesses for the optimal measurement set using one material. Results shown in section \ref{multiple_materials_impact} illustrate the impact of using multiple materials on the condition number, with two distinct cases.

Using multiple piled up materials is beneficial in comparison to the previous setup from~\citep{li_em_2021}.
This can be explained by the much larger size of the research space when multiple materials are allowed, leading to a better optimum than the single material algorithm. However, as shown in figure~\ref{fig:piled_up_stability}, the expected drawback of this larger research space is a worse stability. The reduction of the research space size thus plays a key role in the stability improvement.

With the second version of our algorithm presented in~\ref{fig:algorithm_versions}, where only one material is allowed per measurement, an even greater improvement of results is observed with a highly reduced condition number and a good stability, shown in figures~\ref{fig:results_separation}, \ref{fig:separated_stability}. Even though the research space is smaller than in the previous version of the algorithm (which theoretically includes this one's particular case), the relevant constraints put on the optimizer allow a more thorough exploration, leading to the discovery of a better optimum.

\subsection{Reduction of scatter noise}

Reduction of the measurement scatter noise has proven to be necessary when a challenging high-energy spectrum is reconstructed. Indeed, even though the ill-posedness of the inverse problem is reduced by the search of optimal measurement sets, the condition number remains significant enough to disturb the reconstruction when the measurement noise is too high.

As displayed in figure~\ref{fig:barillet_tot_dir}, when a naive simulation configuration is used (config. 1) the scatter noise tends to become as large as $100 \%$, leading to measurements unusable for reconstruction.

When the measurement slabs are optimally distributed within the experimental circle (config. 2), figure~\ref{fig:hedgehog_tot_dir} shows a substantial reduction of scattering, with unscattered rays noised by an amount of $50 \%$. This scatter noise is however still way too high for the inversion.

Lastly, when a thick lead anti-scatter grid is added behind the measurement slabs (config. 3), the scatter noise is reduced to a negligible amount of around $3 \%$ as shown in figure~\ref{fig:hedgehog_shield_tot_dir}. The lead layer allows the absorption of almost every scattered ray and the detection of the unscattered rays that pass through material slabs parallel to its axis. The scatter noise obtained with this last configuration appears to be small enough for the measurements to be used to solve the inverse problem and reconstruct the spectrum.

\subsection{Schiff spectrum reconstruction}

\subsubsection{Nominal configuration}

As shown in figure~\ref{fig:nominal_reconstruction}, the overall reconstruction accuracy is satisfying in the nominal configuration, but two areas can be distinguished. For high energy ($E > 100$ keV) there is no restriction as the constraint step function is set to unity, and the reconstruction shows great accuracy even for as few as 12 interpolation points. However, in the low energy range, the points are constrained to 0 and are thus not optimized. In practice, this is necessary because of the much lower contribution of low energies in the transmission measurements. Indeed, when dense materials are subjected to an X-ray beam of given spectrum, most of the low-energy photons are absorbed and are not detected at the end. The difficulty to reconstruct the low energy part of spectra is thus an intrinsic issue for the inverse problem at hand. 

\subsubsection{Ablation studies}

Figure~\ref{fig:no_anti-scatter_grid_reconstruction} emphasizes the substantial impact of the presence of an anti-scatter grid in the experimental setup on the reconstruction accuracy. When nothing is done to reduce the scattered rays, the measurements are very noisy, leading inevitably to an inaccurate reconstruction.

Similarly, the influence of using multiple materials is illustrated in figure~\ref{fig:1_mat_reconstruction}. Because only iron is allowed in the first optimization problem, the genetic algorithm cannot converge to a satisfying minimum of the condition number of the system. Thus, even with low noise, the reconstruction is unstable and inaccurate.

Finally, the interest of using the "trust-constr" algorithm for the second optimization problem formulated above is clearly highlighted in figure~\ref{fig:EM_reconstruction}. When the expectation-maximization algorithm is used instead, as it has usually been done in other studies~\cite{li_em_2021}, the quality of the reconstruction is very poor for low and medium energy ranges, and the spectrum peak is not retrieved at the expected energy level.

\section{CONCLUSION} 
In this article, a full spectrum reconstruction pipeline for high-energy X-ray sources was presented. Building on prior work which introduced the optimization of transmission measurements, the present work generalizes this approach to multiple calibration materials, enabling to reach better performance than thickness-only optimization. This work also demonstrated the importance of noise reduction to perform spectrum reconstruction in realistic experimental setups. As such, it showed that the design of an adapted anti-scatter grid is a precious asset to solve the inverse problem and obtain faithful spectrum estimations. Finally, a novel noise-robust reconstruction method was shown to outperform common expectation-maximization approaches, enabling a precise choice of spectrum resolution and a controlled injection of prior knowledge of the X-ray spectrum.

\section*{Conflict of Interest Statement}
The authors declare that the research was conducted in the absence of any commercial or financial relationships that could be construed as a potential conflict of interest.

\section*{Author Contributions}
KG and AF lead the research. AW wrote the code and conducted numerical experiments. AW, KG and AF wrote the article. 

\section*{Funding}
No funding information applicable.

\bibliographystyle{unsrtnat}
\bibliography{spectrum_reconstruction}  

\begin{thebibliography}{15}
\providecommand{\natexlab}[1]{#1}
\providecommand{\url}[1]{\texttt{#1}}
\expandafter\ifx\csname urlstyle\endcsname\relax
  \providecommand{\doi}[1]{doi: #1}\else
  \providecommand{\doi}{doi: \begingroup \urlstyle{rm}\Url}\fi

\bibitem[Sidky et~al.(2004)Sidky, Yu, and Pan]{yaffe_application_2004}
Emil~Y. Sidky, Lifeng Yu, and Xiaochuan Pan.
\newblock Application of expectation maximization to x-ray spectrum estimation
  for medical accelerators from transmission data.
\newblock page 856, San Diego, CA, May 2004.
\newblock \doi{10.1117/12.535989}.
\newblock URL
  \url{http://proceedings.spiedigitallibrary.org/proceeding.aspx?doi=10.1117/12.535989}.

\bibitem[Duan et~al.(2011)Duan, Wang, Yu, Leng, and McCollough]{duan_ct_2011}
Xinhui Duan, Jia Wang, Lifeng Yu, Shuai Leng, and Cynthia~H. McCollough.
\newblock {CT} scanner x-ray spectrum estimation from transmission
  measurements: {CT} scanner x-ray spectrum estimation from transmission
  measurements.
\newblock \emph{Medical Physics}, 38\penalty0 (2):\penalty0 993--997, January
  2011.
\newblock ISSN 00942405.
\newblock \doi{10.1118/1.3547718}.
\newblock URL \url{http://doi.wiley.com/10.1118/1.3547718}.

\bibitem[Wood(2018)]{wood_shot-by-shot_2018}
Wm~M. Wood.
\newblock Shot-by-shot spectrum model for rod-pinch, pulsed radiography
  machines.
\newblock \emph{AIP Advances}, 8\penalty0 (2):\penalty0 025105, February 2018.
\newblock ISSN 2158-3226.
\newblock \doi{10.1063/1.5016299}.
\newblock URL \url{http://aip.scitation.org/doi/10.1063/1.5016299}.

\bibitem[Waggener et~al.(1999)Waggener, Blough, Terry, Chen, Lee, Zhang, and
  McDavid]{waggener_x-ray_1999}
Robert~G. Waggener, Melissa~M. Blough, James~A. Terry, Di~Chen, Nina~E. Lee,
  Sean Zhang, and William~D. McDavid.
\newblock X-ray spectra estimation using attenuation measurements from 25 {kVp}
  to 18 {MV}.
\newblock \emph{Medical Physics}, 26\penalty0 (7):\penalty0 1269--1278, July
  1999.
\newblock ISSN 00942405.
\newblock \doi{10.1118/1.598622}.
\newblock URL \url{http://doi.wiley.com/10.1118/1.598622}.

\bibitem[Armbruster et~al.(2004)Armbruster, Hamilton, and
  Kuehl]{armbruster_spectrum_2004}
Benjamin Armbruster, Russell~J Hamilton, and Arthur~K Kuehl.
\newblock Spectrum reconstruction from dose measurements as a linear inverse
  problem.
\newblock \emph{Physics in Medicine and Biology}, 49\penalty0 (22):\penalty0
  5087--5099, November 2004.
\newblock ISSN 0031-9155, 1361-6560.
\newblock \doi{10.1088/0031-9155/49/22/005}.
\newblock URL
  \url{https://iopscience.iop.org/article/10.1088/0031-9155/49/22/005}.

\bibitem[Paniak and Charland(2005)]{paniak_enhanced_2005}
L~D Paniak and P~M Charland.
\newblock Enhanced bremsstrahlung spectrum reconstruction from depth–dose
  gradients.
\newblock \emph{Physics in Medicine and Biology}, 50\penalty0 (14):\penalty0
  3245--3261, July 2005.
\newblock ISSN 0031-9155, 1361-6560.
\newblock \doi{10.1088/0031-9155/50/14/004}.
\newblock URL
  \url{https://iopscience.iop.org/article/10.1088/0031-9155/50/14/004}.

\bibitem[Sidky et~al.(2005)Sidky, Yu, Pan, Zou, and Vannier]{sidky_robust_2005}
Emil~Y. Sidky, Lifeng Yu, Xiaochuan Pan, Yu~Zou, and Michael Vannier.
\newblock A robust method of x-ray source spectrum estimation from transmission
  measurements: {Demonstrated} on computer simulated, scatter-free transmission
  data.
\newblock \emph{Journal of Applied Physics}, 97\penalty0 (12), 2005.
\newblock ISSN 0021-8979.
\newblock \doi{10.1063/1.1928312}.
\newblock URL
  \url{http://www.scopus.com/inward/record.url?scp=21644449154&partnerID=8YFLogxK}.

\bibitem[Zhao et~al.(2015)Zhao, Niu, Schafer, and Royalty]{zhao_indirect_2015}
Wei Zhao, Kai Niu, Sebastian Schafer, and Kevin Royalty.
\newblock An indirect transmission measurement-based spectrum estimation method
  for computed tomography.
\newblock \emph{Physics in Medicine and Biology}, 60\penalty0 (1):\penalty0
  339--357, January 2015.
\newblock ISSN 0031-9155, 1361-6560.
\newblock \doi{10.1088/0031-9155/60/1/339}.
\newblock URL
  \url{https://iopscience.iop.org/article/10.1088/0031-9155/60/1/339}.

\bibitem[FitzGerald et~al.(2021)FitzGerald, Araujo, Wu, and
  De~Man]{fitzgerald_semiempirical_2021}
Paul FitzGerald, Stephen Araujo, Mingye Wu, and Bruno De~Man.
\newblock Semiempirical, parameterized spectrum estimation for x‐ray computed
  tomography.
\newblock \emph{Medical Physics}, 48\penalty0 (5):\penalty0 2199--2213, May
  2021.
\newblock ISSN 0094-2405, 2473-4209.
\newblock \doi{10.1002/mp.14715}.
\newblock URL \url{https://onlinelibrary.wiley.com/doi/10.1002/mp.14715}.

\bibitem[Li et~al.(2021)Li, Fan, Cong, and Wang]{li_em_2021}
Mengzhou Li, Feng-Lei Fan, Wenxiang Cong, and Ge~Wang.
\newblock {EM} {Estimation} of the {X}-{Ray} {Spectrum} {With} a {Genetically}
  {Optimized} {Step}-{Wedge} {Phantom}.
\newblock \emph{Frontiers in Physics}, 9:\penalty0 678171, May 2021.
\newblock ISSN 2296-424X.
\newblock \doi{10.3389/fphy.2021.678171}.
\newblock URL
  \url{https://www.frontiersin.org/articles/10.3389/fphy.2021.678171/full}.

\bibitem[Schiff(1951)]{schiff_energy-angle_1951}
L.~I. Schiff.
\newblock Energy-{Angle} {Distribution} of {Thin} {Target} {Bremsstrahlung}.
\newblock \emph{Physical Review}, 83\penalty0 (2):\penalty0 252--253, July
  1951.
\newblock ISSN 0031-899X.
\newblock \doi{10.1103/PhysRev.83.252}.
\newblock URL \url{https://link.aps.org/doi/10.1103/PhysRev.83.252}.

\bibitem[Estre et~al.(2013)Estre, Eck, Pettier, Payan, Roure, and
  Simon]{estre_high-energy_2013}
Nicolas Estre, Daniel Eck, Jean-Luc Pettier, Emmanuel Payan, Christophe Roure,
  and Eric Simon.
\newblock High-energy {X}-ray imaging applied to non destructive
  characterization of large nuclear waste drums.
\newblock In \emph{2013 3rd {International} {Conference} on {Advancements} in
  {Nuclear} {Instrumentation}, {Measurement} {Methods} and their {Applications}
  ({ANIMMA})}, pages 1--6, June 2013.
\newblock \doi{10.1109/ANIMMA.2013.6727987}.

\bibitem[Storn and Price(1997)]{storn1997differential}
Rainer Storn and Kenneth Price.
\newblock Differential evolution--a simple and efficient heuristic for global
  optimization over continuous spaces.
\newblock \emph{Journal of global optimization}, 11:\penalty0 341--359, 1997.

\bibitem[Akima(1970)]{akima1970new}
Hiroshi Akima.
\newblock A new method of interpolation and smooth curve fitting based on local
  procedures.
\newblock \emph{Journal of the ACM (JACM)}, 17\penalty0 (4):\penalty0 589--602,
  1970.

\bibitem[Conn et~al.(2000)Conn, Gould, and Toint]{conn2000trust}
Andrew~R Conn, Nicholas~IM Gould, and Philippe~L Toint.
\newblock \emph{Trust region methods}.
\newblock SIAM, 2000.

\end{thebibliography}






\end{document}